\documentclass[aps,prl,preprint,floatfix,superscriptaddress,showpacs]{revtex4}

\usepackage{graphicx}
\usepackage{dcolumn}
\usepackage{ulem}
\usepackage{soul}

\begin{document}

\title{PAMELA results on the cosmic-ray antiproton flux}

\author{O. Adriani}
\affiliation{University of Florence, Department of Physics,  
 I-50019 Sesto Fiorentino, Florence, Italy}
\affiliation{INFN, Sezione di Florence,  
 I-50019 Sesto Fiorentino, Florence, Italy}
\author{G. C. Barbarino}
\affiliation{University of Naples ``Federico II'', Department of
Physics, I-80126 Naples, Italy}
\affiliation{INFN, Sezione di Naples,  I-80126 Naples, Italy}
\author{G. A. Bazilevskaya}
\affiliation{Lebedev Physical Institute, RU-119991
Moscow, Russia}
\author{R. Bellotti}
\affiliation{University of Bari, Department of Physics, I-70126 Bari, Italy}
\affiliation{INFN, Sezione di Bari, I-70126 Bari, Italy}
\author{M. Boezio}
\affiliation{INFN, Sezione di Trieste, I-34149
Trieste, Italy}
\author{E. A. Bogomolov}
\affiliation{Ioffe Physical Technical Institute,  RU-194021 St. 
Petersburg, Russia}
\author{L. Bonechi}
\affiliation{University of Florence, Department of Physics,  
 I-50019 Sesto Fiorentino, Florence, Italy}
\affiliation{INFN, Sezione di Florence,  
 I-50019 Sesto Fiorentino, Florence, Italy}
\author{M. Bongi}
\affiliation{INFN, Sezione di Florence,  
 I-50019 Sesto Fiorentino, Florence, Italy}
\author{V. Bonvicini}
\affiliation{INFN, Sezione di Trieste,  I-34149
Trieste, Italy}
\author{S. Borisov}
\affiliation{INFN, Sezione di Rome ``Tor Vergata'', I-00133 Rome, Italy}
\affiliation{University of Rome ``Tor Vergata'', Department of
Physics,  I-00133 Rome, Italy} 
\affiliation{Moscow Engineering and Physics Institute,  RU-11540
Moscow, Russia}  
\author{S. Bottai}
\affiliation{INFN, Sezione di Florence,  
 I-50019 Sesto Fiorentino, Florence, Italy}
\author{A. Bruno}
\affiliation{University of Bari, Department of Physics, I-70126 Bari,
Italy} 
\affiliation{INFN, Sezione di Bari, I-70126 Bari, Italy}
\author{F. Cafagna}
\affiliation{INFN, Sezione di Bari, I-70126 Bari, Italy}
\author{D. Campana}
\affiliation{INFN, Sezione di Naples,  I-80126 Naples, Italy}
\author{R. Carbone}
\affiliation{INFN, Sezione di Naples,  I-80126 Naples, Italy}
\affiliation{University of Rome ``Tor Vergata'', Department of
Physics,  I-00133 Rome, Italy} 
\author{P. Carlson}
\affiliation{KTH, Department of Physics, and the Oskar Klein Centre for
Cosmoparticle Physics, AlbaNova University Centre, SE-10691 Stockholm,
Sweden}
\author{M. Casolino}
\affiliation{INFN, Sezione di Rome ``Tor Vergata'', I-00133 Rome, Italy}
\author{G. Castellini}
\affiliation{ IFAC,  I-50019 Sesto Fiorentino,
Florence, Italy}
\author{L. Consiglio}
\affiliation{INFN, Sezione di Naples,  I-80126 Naples, Italy}
\author{M. P. De Pascale}
\affiliation{INFN, Sezione di Rome ``Tor Vergata'', I-00133 Rome, Italy}
\affiliation{University of Rome ``Tor Vergata'', Department of
Physics,  I-00133 Rome, Italy} 
\author{C. De Santis}
\affiliation{INFN, Sezione di Rome ``Tor Vergata'', I-00133 Rome, Italy}
\author{N. De Simone}
\affiliation{INFN, Sezione di Rome ``Tor Vergata'', I-00133 Rome, Italy}
\affiliation{University of Rome ``Tor Vergata'', Department of
Physics,  I-00133 Rome, Italy} 
\author{V. Di Felice}
\affiliation{INFN, Sezione di Rome ``Tor Vergata'', I-00133 Rome, Italy}
\affiliation{University of Rome ``Tor Vergata'', Department of
Physics,  I-00133 Rome, Italy} 
\author{A. M. Galper}
\affiliation{Moscow Engineering and Physics Institute,  RU-11540
Moscow, Russia}  
\author{W. Gillard}
\affiliation{KTH, Department of Physics, and the Oskar Klein Centre for
Cosmoparticle Physics, AlbaNova University Centre, SE-10691 Stockholm,
Sweden}
\author{L. Grishantseva}
\affiliation{Moscow Engineering and Physics Institute,  RU-11540
Moscow, Russia}  
\author{P. Hofverberg}
\affiliation{KTH, Department of Physics, and the Oskar Klein Centre for
Cosmoparticle Physics, AlbaNova University Centre, SE-10691 Stockholm,
Sweden}
\author{G. Jerse}
\affiliation{INFN, Sezione di Trieste, I-34149
Trieste, Italy}
\affiliation{University of Trieste, Department of Physics, 
I-34147 Trieste, Italy}
\author{A. V. Karelin}
\affiliation{Moscow Engineering and Physics Institute,  RU-11540
Moscow, Russia}
\author{S. V. Koldashov}
\affiliation{Moscow Engineering and Physics Institute,  RU-11540
Moscow, Russia}  
\author{S. Y. Krutkov}
\affiliation{Ioffe Physical Technical Institute,  RU-194021 St. 
Petersburg, Russia}
\author{A. N. Kvashnin}
\affiliation{Lebedev Physical Institute, RU-119991
Moscow, Russia}
\author{A. Leonov}
\affiliation{Moscow Engineering and Physics Institute,  RU-11540
Moscow, Russia}  
\author{V. Malvezzi}
\affiliation{INFN, Sezione di Rome ``Tor Vergata'', I-00133 Rome, Italy}
\author{L. Marcelli}
\affiliation{INFN, Sezione di Rome ``Tor Vergata'', I-00133 Rome, Italy}
\author{A. G. Mayorov}
\affiliation{Moscow Engineering and Physics Institute,  RU-11540
Moscow, Russia}
\author{W. Menn}
\affiliation{Universit\"{a}t Siegen, Department of Physics,
D-57068 Siegen, Germany}
\author{V. V. Mikhailov}
\affiliation{Moscow Engineering and Physics Institute,  RU-11540
Moscow, Russia}  
\author{E. Mocchiutti}
\affiliation{INFN, Sezione di Trieste,  I-34149
Trieste, Italy}
\author{A. Monaco}
\affiliation{University of Bari, Department of Physics, I-70126 Bari, Italy}
\affiliation{INFN, Sezione di Bari, I-70126 Bari, Italy}
\author{N. Mori}
\affiliation{INFN, Sezione di Florence,  
 I-50019 Sesto Fiorentino, Florence, Italy}
\author{N. Nikonov}
\affiliation{Ioffe Physical Technical Institute,  RU-194021 St. 
Petersburg, Russia}
\affiliation{INFN, Sezione di Rome ``Tor Vergata'', I-00133 Rome, Italy}
\affiliation{University of Rome ``Tor Vergata'', Department of
Physics,  I-00133 Rome, Italy} 
\author{G. Osteria}
\affiliation{INFN, Sezione di Naples,  I-80126 Naples, Italy}
\author{P. Papini}
\affiliation{INFN, Sezione di Florence,  
 I-50019 Sesto Fiorentino, Florence, Italy}
\author{M. Pearce}
\affiliation{KTH, Department of Physics, and the Oskar Klein Centre for
Cosmoparticle Physics, AlbaNova University Centre, SE-10691 Stockholm,
Sweden}
\author{P. Picozza}
\affiliation{INFN, Sezione di Rome ``Tor Vergata'', I-00133 Rome, Italy}
\affiliation{University of Rome ``Tor Vergata'', Department of
Physics,  I-00133 Rome, Italy} 
\author{C. Pizzolotto}
\affiliation{INFN, Sezione di Trieste, I-34149
Trieste, Italy}
\author{M. Ricci}
\affiliation{INFN, Laboratori Nazionali di Frascati, Via Enrico Fermi 40,
I-00044 Frascati, Italy}
\author{S. B. Ricciarini}
\affiliation{INFN, Sezione di Florence, 
 I-50019 Sesto Fiorentino, Florence, Italy}
\author{L. Rossetto}
\affiliation{KTH, Department of Physics, and the Oskar Klein Centre for
Cosmoparticle Physics, AlbaNova University Centre, SE-10691 Stockholm,
Sweden}
\author{M. Simon}
\affiliation{Universit\"{a}t Siegen, Department of Physics,
D-57068 Siegen, Germany}
\author{R. Sparvoli}
\affiliation{INFN, Sezione di Rome ``Tor Vergata'', I-00133 Rome, Italy}
\affiliation{University of Rome ``Tor Vergata'', Department of
Physics,  I-00133 Rome, Italy} 
\author{P. Spillantini}
\affiliation{University of Florence, Department of Physics,  
 I-50019 Sesto Fiorentino, Florence, Italy}
\affiliation{INFN, Sezione di Florence,  
 I-50019 Sesto Fiorentino, Florence, Italy}
\author{Y. I. Stozhkov}
\affiliation{Lebedev Physical Institute, RU-119991
Moscow, Russia}
\author{A. Vacchi}
\affiliation{INFN, Sezione di Trieste,  I-34149
Trieste, Italy}
\author{E. Vannuccini}
\affiliation{INFN, Sezione di Florence, 
 I-50019 Sesto Fiorentino, Florence, Italy}
\author{G. Vasilyev}
\affiliation{Ioffe Physical Technical Institute, RU-194021 St. 
Petersburg, Russia}
\author{S. A. Voronov}
\affiliation{Moscow Engineering and Physics Institute,  RU-11540
Moscow, Russia}
\author{J. Wu}
\altaffiliation[On leave from ]{School of Mathematics and Physics,
China University of Geosciences, CN-430074 Wuhan, China}
\affiliation{KTH, Department of Physics, and the Oskar Klein Centre for
Cosmoparticle Physics, AlbaNova University Centre, SE-10691 Stockholm,
Sweden}
\author{Y. T. Yurkin}
\affiliation{Moscow Engineering and Physics Institute,  RU-11540
Moscow, Russia}  
\author{G. Zampa}
\affiliation{INFN, Sezione di Trieste,  I-34149
Trieste, Italy}
\author{N. Zampa}
\affiliation{INFN, Sezione di Trieste,  I-34149
Trieste, Italy}
\author{V. G. Zverev}
\affiliation{Moscow Engineering and Physics Institute,  RU-11540
Moscow, Russia}

\date{\today}

\begin{abstract}
The satellite-borne experiment PAMELA has been used to make a new
measurement of the cosmic-ray antiproton flux and the antiproton-to-proton
flux ratio which extends previously published measurements down to 60 MeV
and up to 180 GeV in kinetic energy. During 850 days of data acquisition
approximately 1500 antiprotons were observed. The measurements are
consistent with purely secondary production of antiprotons in the galaxy.
More precise secondary production models are required for a complete
interpretation of the results.

\end{abstract}


\maketitle


Antiprotons and positrons are a small but not negligible component of
the cosmic radiation. They can be 
produced in the interactions between
cosmic-ray nuclei and the interstellar matter. Detailed measurements
of the cosmic-ray antiproton energy spectrum therefore provide important
information concerning the origin and propagation of cosmic-rays. 
Exotic sources of primary antiprotons such as 
the annihilation of dark matter
particles~\cite{jun96,ber00,ber05} and  
the evaporation
of primordial 
black holes~\cite{haw74,kir81} can also be probed.
The theoretical energy spectrum of secondary antiprotons has a
distinct peak around 2~GeV and rapidly decreases towards lower
energies due to the  
kinematic constraints on the antiproton production. At higher energies
the spectrum is slightly steeper than that of the parent protons
(e.g. see \cite{sim98}), 
which 
results in a slight decrease of the antiproton-to-proton flux ratio. 

Since July 2006, PAMELA (a Payload for
Antimatter Matter Exploration and Light-nuclei Astrophysics) is
measuring the antiparticle component of the cosmic radiation. 
A previous PAMELA measurement of the antiproton-to-proton flux ratio
between 1.5 and 100 GeV~\cite{adr09a}, was found to follow the 
expectation from secondary production calculations. However, the
positron fraction~\cite{adr09b,adr10} measured in the same energy
range showed a clear deviation from secondary production models. In
order to explain these results both astrophysical objects
(e.g. pulsars) 
and dark matter have been proposed as positron sources
(e.g.~\cite{boe09}). A contribution from 
pulsars would naturally increase the
positron and electron abundances without affecting the antiproton
component. Other astrophysical
models \cite{bla09b} have been proposed to explain the PAMELA
positron results but 
produce an increase in the antiproton component at very high energies
($\geq$100 GeV). 
A dark matter contribution may
require pure leptonic annihilation 
channels, 
e.g.~\cite{cir08}, or the introduction of a new dark sector
of forces,
e.g.~\cite{cho08}. In \cite{kan09} 
it is noted that any signal in the antiproton energy spectrum may be
hidden due to  
incomplete modelling of secondary production and cosmic-ray
propagation. A detailed measurement of the antiproton energy
spectrum over a large energy range is therefore of great interest.

The PAMELA experiment \cite{pic07,boe09} comprises 
(from top to bottom): a time of flight system,
a magnetic spectrometer with silicon tracker planes,
an anticoincidence system,
an electromagnetic imaging calorimeter,
a shower tail catcher scintillator and 
a neutron detector. These components
are housed inside a
pressurized container attached
to the Russian Resurs-DK1 satellite, which was launched on
June 15$^{{\rm th}}$ 2006. The orbit is elliptical
and \mbox{semi-polar}, with an inclination of 70.0$^\circ$ and an
altitude varying between 350~km and 610~km. 

We report on the cosmic-ray antiproton flux over the widest energy range ever
achieved: 60 MeV to 180 GeV. We also confirm and extend the previously
published PAMELA 
antiproton-to-proton flux ratio measurement \cite{adr09a} to the same
energy range. 
Data were acquired from July 2006 to December 2008 (850 days),
corresponding to $>10^9$ triggers. 
Triggered events were selected for analysis if the reconstructed
rigidity exceeded the 
vertical geomagnetic cut-off (estimated using the satellite orbital
information) by a factor of 1.3. Downward-going charge-one particles
were selected using the 
time-of-flight and spectrometer data. 
Time-of-flight information was also used to select low velocity
(anti)protons while electrons were rejected using the electromagnetic
calorimeter information, as described in \cite{adr09a}. The remaining
electron 
contamination was estimated to be negligible while contamination from
locally produced pions was found to be about 10\% between 1 and 3 GV/c and
negligible at lower and higher rigidities~\cite{adr09a,bru08}.

The highest energy at which antiprotons can be unambiguously measured
by PAMELA is determined by the contamination of
``spillover'' protons which are reconstructed with an incorrect
sign of curvature either due to the finite
spectrometer resolution or scattering in the spectrometer planes.
To reduce this contamination, strict requirements were applied on the
quality of the tracks reconstructed in the spectrometer.  
For example, tracks accompanied  by $\delta$-ray emission were
discarded to avoid poorly reconstructed coordinates on the silicon planes of the
spectrometer. 
For each track the maximum detectable rigidity (MDR) was evaluated on an 
event-by-event basis by propagating the estimated coordinate errors and taking
into account the track topology. The MDR was required to be 6 times larger than the measured
rigidity. This allowed the antiproton measurement to be extended up to
180~GV/c with acceptable contamination from spillover protons. 
The contamination was estimated using the GPAMELA detector simulation
which is based on the GEANT3 package \cite{bru94}.  
The simulation contains an accurate representation of the geometry and
performance of the PAMELA detectors. For the spectrometer~\cite{str06} the
measured noise of each silicon plane and  
performance variations over the duration of the measurement were
accounted for.  
The simulation code was validated by comparing the distributions of several
significant variables (e.g. coordinate residuals, $\chi^2$ and
the covariance matrix from the track fitting) with those obtained from real data. 
The high-energy region of the deflection distribution was studied before applying the MDR selection and  
agreement within 20\% was found between data and simulation. This difference was taken as a systematic 
uncertainty on the spillover contamination which was estimated to be $\simeq 30\%$ for the rigidity
interval 100-180~GV/c.

The efficiencies were carefully studied using both experimental and
simulated data~\cite{hof08,bru08,wu10}. 
The time dependence of the detector performance (and therefore also efficiency) was studied using proton samples collected during 2 month long periods.
The average global
selection efficiency was measured to be $\simeq 30\%$.
The number of (anti)protons rejected by the selection criteria 
due to interactions and energy loss within the detector systems was estimated using the
simulation. The number of antiprotons lost due to
this selection is energy dependent and varies from $\simeq 10\%$ below 1 GeV to
$\simeq 6\%$ above 50 GeV. 
The antiproton flux was obtained by considering the geometrical factor (estimated both analytically and with simulations)  
and the total live time which is provided by an on-board clock that times the
periods during which the apparatus is waiting for a trigger.

The energy-binned antiproton fluxes and antiproton-to-proton flux ratios 
are given in Table~\ref{t:pbar}. 
The spectrometer resolution has not been unfolded and a systematic
uncertainty is included to account for this.
\begin{table}
\caption{Summary of antiproton 
results. Antiproton fluxes ($\times 10^{-3}$ particles/(m$^{2}$ sr s
GeV)) and antiproton-to-proton flux ratios ($\times
10^{-5}$). The upper limits are 90\% confidence levels. The first and
second errors represent the statistical and 
systematic uncertainties, respectively. \label{t:pbar}}   
\begin{ruledtabular}
\begin{tabular}{ccccc}
Rigidity & Mean Kinetic & Observed & Flux & \(
\frac{\mbox{$\overline{{\rm p}}$ }}{{\mbox p}} \)  
\\ 
at the & Energy at & number of & at top of & at top of \\  
spectrometer & top of & events \mbox{$\overline{{\rm p}}$ } & payload
 & payload \\
GV/c & payload GeV & 
 &  & 
 \\ \hline
    0.35 -   0.50 &   0.09 &    0 & $  < 6.4 $ & $  < 0.73 $ \\
    0.50 -   1.01 &   0.28 &    7 & $   6.7 \pm    2.7 \pm    0.2$ & $
0.48 \pm   0.18 \pm 0.01$ \\
    1.01 -   1.34 &   0.56 &   15 & $  15.3 ^{+   7.5}_{-
3.7} \pm 0.9$ & $  0.99 ^{+  0.31}_{-  0.26} \pm 0.07 $ \\
    1.34 -   1.63 &   0.81 &   19 & $  17.2 ^{+   7.4}_{-
3.9} \pm 1.1 $ & $  1.33 ^{+  0.38}_{-  0.33} \pm 0.10 $ \\
    1.63 -   1.93 &   1.07 &   32 & $  21.4 ^{+   6.8}_{-
3.9} \pm 1.3 $ & $  2.04 \pm 0.44 \pm 0.15 $ \\
    1.93 -   2.23 &   1.34 &   39 & $  24.5 ^{+   7.2}_{-
4.3} \pm 1.5 $ & $  2.78 \pm 0.54 \pm 0.20 $ \\
    2.23 -   2.58 &   1.61 &   49 & $  20.5 \pm    3.2 \pm    1.2$ & $  3.43 \pm   0.49 \pm   0.24$ \\
    2.58 -   2.99 &   2.03 &   78 & $  27.1 \pm    3.3 \pm    1.6$ & $  5.44 \pm   0.62 \pm   0.39$ \\
    2.99 -   3.45 &   2.42 &   79 & $  21.9 \pm    2.6 \pm    1.3$ & $  6.10 \pm   0.68 \pm   0.43$ \\
    3.45 -   3.99 &   2.90 &   96 & $  22.7 \pm    2.5 \pm    1.3$ & $  7.78 \pm   0.79 \pm   0.55$ \\
    3.99 -   4.62 &   3.47 &  103 & $  17.8 \pm    1.9 \pm    1.0$ & $  9.15 \pm   0.89 \pm   0.65$ \\
    4.62 -   5.36 &   4.14 &  109 & $  15.7 \pm    1.6 \pm    0.9$ & $  10.7 \pm    1.0 \pm    0.8$ \\
    5.36 -   6.23 &   4.93 &  110 & $  11.1 \pm    1.1 \pm    0.7$ & $  12.0 \pm    1.1 \pm    0.9$ \\
     6.2 -    7.3 &    5.9 &  106 & $  8.31 \pm   0.86 \pm   0.49$ & $  12.5 \pm    1.2 \pm    0.9$ \\
     7.3 -    8.5 &    7.0 &   87 & $  5.56 \pm   0.64 \pm   0.33$ & $  12.2 \pm    1.3 \pm    0.9$ \\
     8.5 -   10.1 &    8.4 &   98 & $  5.16 \pm   0.57 \pm   0.30$ & $  15.6 \pm    1.6 \pm    1.1$ \\
    10.1 -   12.0 &   10.1 &  108 & $  3.70 \pm   0.38 \pm   0.22$ & $  20.8 \pm    1.9 \pm    1.5$ \\
    12.0 -   14.6 &   12.3 &   82 & $  2.12 \pm   0.26 \pm   0.12$ & $  16.1 \pm    1.8 \pm    1.1$ \\
    14.6 -   18.1 &   15.3 &   64 & $  1.39 \pm   0.19 \pm   0.08$ & $  20.7 \pm    2.4 \pm    1.5$ \\
    18.1 -   23.3 &   19.6 &   56 & $  0.67 \pm   0.10 \pm   0.04$ & $  17.4 \pm    2.2 \pm    1.2$ \\
    23.3 -   31.7 &   26.2 &   42 & $ 0.251 \pm  0.041 \pm  0.015$ & $  17.1 \pm    2.5 \pm    1.2$ \\
    31.7 -   48.5 &   38.0 &   36 & $ 0.127 \pm  0.023 \pm  0.007$ & $  18.3 \pm    3.0 \pm    1.3$ \\
    48.5 -  100.0 &   67.4 &   22 & $0.0228 \pm 0.0072 \pm 0.0008$ & $  17.7 \pm    4.8 \pm    0.8$ \\
   100.0 -  180.0 &  128.9 &    3 & $0.0036 ^{+0.0057}_{-0.0020} \pm 0.0002$ & $   14 ^{+   16 }_{-   10} \pm    1$ \\

\end{tabular}
\end{ruledtabular}
\end{table}
Contamination from pions and spillover protons has been subtracted from the
results. 
The first and second errors in the table represent the statistical and
systematic uncertainties, respectively. The total systematic
uncertainty was obtained quadratically summing the various systematic
errors considered: acceptance, contamination, efficiency
estimation, energy losses, interactions and spectrum unfolding.

\begin{figure}[h]
\includegraphics[width=25pc]{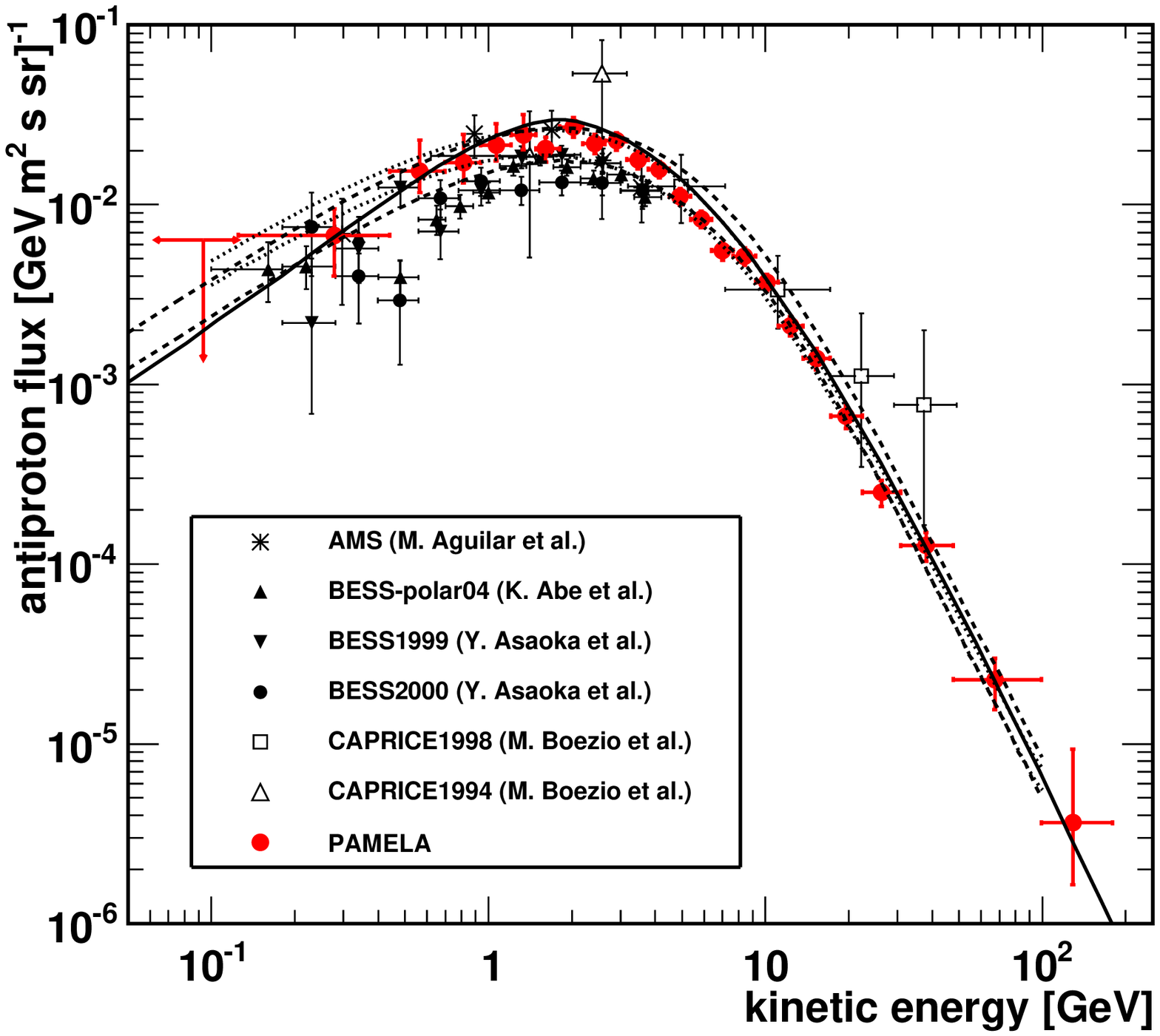}\hspace{2pc}
\caption{The antiproton energy spectrum at the top of
the payload obtained in 
this work compared with contemporary
measurements~\protect\cite{boe97,boe01,asa02,abe08,agu02} and  
theoretical calculations for a pure secondary
production of antiprotons during the propagation of cosmic rays in the
galaxy. 
The dotted and dashed lines indicate 
the upper and lower limits calculated by 
\citet{don01} for different diffusion models, including uncertainties on
propagation parameters and antiproton 
production cross-sections, respectively. 
The solid line
shows the calculation by \citet{ptu06} for
the case of a Plain Diffusion model. 
\label{flux}}   
\end{figure}

\begin{figure}[h]
\includegraphics[width=25pc]{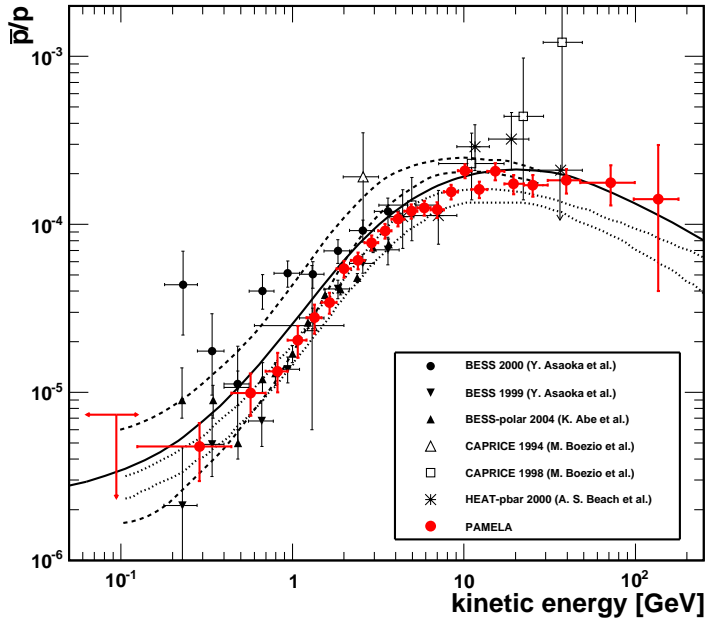}\hspace{2pc}
\caption{The antiproton-to-proton flux ratio at the top of
the payload obtained in 
this work compared with contemporary
measurements~\protect\cite{boe97,boe01,asa02,abe08,bea01} and 
theoretical calculations for a pure secondary
production of antiprotons during the propagation of cosmic rays in the
galaxy. 
The dashed lines
show the upper and lower limits calculated by \citet{sim98} for the
Leaky Box Model, while the dotted
lines show the limits from \citet{don09} for a Diffusion
Reacceleration with Convection model. 
The solid line
shows the calculation by \citet{ptu06} for
the case of a Plain Diffusion model. 
\label{ratio1}}   
\end{figure}

Figure~\ref{flux} shows the antiproton energy spectrum and
Figure~\ref{ratio1} shows the antiproton-to-proton flux ratio measured 
by PAMELA 
along with other recent experimental
data~\cite{boe97,boe01,asa02,abe08,agu02,bea01} and 
theoretical calculations assuming pure secondary production 
of antiprotons during the propagation of cosmic rays in the
galaxy. The curves were calculated for solar minimum,
which is appropriate for the PAMELA data taking period,  
using the force field approximation \cite{gle68}, 
\footnote{While more precise models
of solar modulation, 
accounting for effects such as sign-charge dependence of the
modulation, exist (e.g.~\cite{bie99,lag04}), the force field model is a
simple approach that provides a reasonably good approximation 
of the solar modulation above 1-2 GeV.}.

\begin{figure}[h]
\includegraphics[width=25pc]{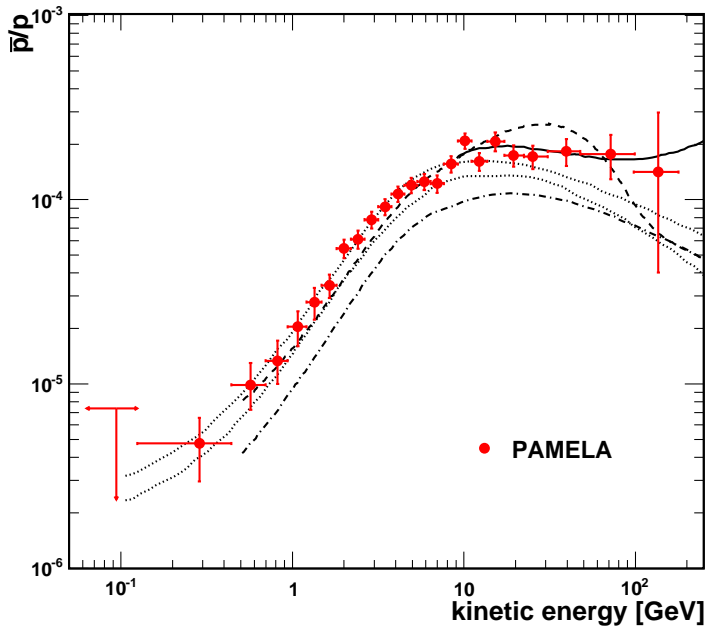}\hspace{2pc}
\caption{The antiproton-to-proton flux ratio at the top of
the payload obtained in 
this work compared with
theoretical calculations. The dotted lines show the upper and lower
limits calculated 
for a pure secondary
production of antiprotons during the propagation of cosmic rays in the
galaxy by \citet{don09} for a Diffusion
Reacceleration with Convection model. 
The dashed line is a calculation by~\citet{kan09} including both a
primary antiproton component 
from annihilation of 180 GeV wino-like neutralinos and
secondary antiprotons (dashed-dotted line for the secondary component).
The solid line show the calculation by~\citet{bla09a} for secondary
antiprotons including an 
additional 
antiproton component produced and accelerated at cosmic-ray sources.
\label{ratio2}}   
\end{figure}

The PAMELA results reproduce the expected peak around 2~GeV in the
antiproton flux and are in overall 
agreement with pure secondary calculations. The experimental
uncertainties are smaller than the spread in the different theoretical 
curves and, therefore, provide important constrains on parameters
relevant for secondary production calculations. For example, the
antiproton flux bands from~\citet{don01} presented in Figure~\ref{flux}  
show 
uncertainties on the propagation parameters (dotted lines) and antiproton
production cross-sections (dashed lines) and indicate larger
uncertainties than those present in the PAMELA measurements. 
Figure~\ref{ratio2} shows
the PAMELA antiproton-to-proton flux ratio compared with 
a calculation~\cite{kan09} (dashed line) including both a
primary antiproton component 
from the annihilation of 180 GeV wino-like neutralinos and
secondary antiprotons. 
This model, based on the non-thermal production of dark matter in the early
universe, was proposed to explain the high-energy rise in the PAMELA positron
fraction~\cite{adr09b}. As shown by the dashed
line in Figure~\ref{ratio2}, a reasonable choice of
GALPROP~\cite{str98} propagation parameters (dashed-dotted 
line) allows a good
description of PAMELA antiproton data with the inclusion of the wino-annihilation signal. 
Given current uncertainties on propagation parameters, this primary component cannot be ruled out. 
It has also been suggested that the PAMELA positron data can be explained without invoking a primary component. 
This is possible if secondary production takes place in the same region where cosmic rays are being
accelerated~\cite{bla09b}. An increase in 
the antiproton~\cite{bla09a} and secondary nuclei
abundances~\cite{mer09} are also predicted in this model.  
The solid line in Figure~\ref{ratio2} shows the prediction for the
high-energy antiproton-to-proton flux ratio.  
While this theoretical prediction is in good agreement with the PAMELA
data, in this energy region it does not differ 
significantly from the expectation for standard secondary production models. 
Comparisons with experimental secondary cosmic-ray nuclei data are
needed along with  
higher energy antiproton measurements. New data on the boron-to-carbon
ratio measured by PAMELA will soon become available, while  
the antiproton spectrum is likely to be probed at higher energies by 
AMS-02 experiment~\cite{bat05} which will soon be placed on the International
Space Station.

We have measured the antiproton energy spectrum and the antiproton-to-proton
flux ratio over the 
most extended energy range ever achieved and with no atmospheric
overburden. Our results are consistent with 
pure secondary production of antiprotons during the propagation of
cosmic rays in the galaxy.
We note that the quality of our data surpasses the current precision 
of the theoretical modeling of the cosmic-ray
acceleration and propagation mechanisms. Improved models are needed to
allow the full significance of these experimental results to be understood.

\begin{acknowledgments}
We acknowledge support from The Italian Space Agency 
(ASI), Deutsches Zentrum f\"{u}r Luft- und Raumfahrt (DLR), The
Swedish National Space  
Board, The Swedish Research Council, The Russian Space Agency
(Roscosmos) and The
Russian Foundation for Basic Research.  
\end{acknowledgments}

\bibliography{pamela_pbarflux_new}

\end{document}